\documentclass{aa}  

\usepackage{graphicx}
\usepackage{txfonts}
\begin{document}

   \title{Deuterated polycyclic aromatic hydrocarbons: Revisited}

   \author{K.D. Doney\inst{1}, A. Candian\inst{1}, T. Mori\inst{2}, T. Onaka\inst{2}, and A.G.G.M. Tielens\inst{1}
          }
   \institute{Leiden Observatory, University of Leiden,
              PO Box 9513, 2300 RA, The Netherlands\\
              \email{doney@strw.leidenuniv.nl}
              \and
              Department of Astronomy, Graduate School of Science, The University of Tokyo,
              7-3-1 Hongo, Bunkyo-ku, Tokyo 113-0033, Japan
             }

 \date{9 December 2015}

 
  \abstract
   {}
   {The amount of deuterium locked up in polycyclic aromatic hydrocarbons (PAHs) has to date been an uncertain value. We present a near-infrared (NIR) spectroscopic survey of H{\sc ii} regions in the Milky Way, Large Magellanic Cloud (LMC), and Small Magellanic Cloud (SMC) obtained with AKARI, which aims to search for features indicative of deuterated PAHs (PAD or D$_n$-PAH) to better constrain the D/H ratio of PAHs.}
   {Fifty-three H{\sc ii} regions were observed in the NIR (2.5-5 $\mu$m), using the Infrared Camera (IRC) on board the AKARI satellite. Through comparison of the observed spectra with a theoretical model of deuterated PAH vibrational modes, the aromatic and (a)symmetric aliphatic C-D stretch modes were identified.}
   {We see emission features  between 4.4-4.8 $\mu$m, which 
could be unambiguously attributed to deuterated PAHs in only six of the observed sources, all of which are located in the Milky Way. In all cases, the aromatic C-D stretching feature is weaker than the aliphatic C-D stretching feature, and, in the case of M17b, this feature is not observed at all. Based on the weak or absent PAD features in most of the observed spectra, it is suggested that the mechanism for PAH deuteration in the ISM is uncommon.}
   {}

   \keywords{astrochemistry - ISM: H{\sc ii} regions - infrared:ISM
               }

   \maketitle
\section{Introduction}   
 All deuterium (D; heavy hydrogen) was formed during the Big Bang and has subsequently been depleted through the process of astration, \textit{i.e.,} destruction by nuclear reactions in the interior of stars \citep{Epstein1976,Linsky2006}. As a result, the amount of deuterium in a galaxy, either as a free atom in the gas phase or locked up in molecules or grains, provides a direct measure of cosmic nucleosynthesis and is related to the chemical evolution of the galaxy itself. Of particular interest is the incorporation of deuterium in polycyclic aromatic hydrocarbon (PAH) molecules.
These molecules are ubiquitous and abundant in the interstellar medium (ISM); their UV/visible-pumped fluorescence is almost universally believed to give rise to aromatic infrared bands (AIBs). These bands dominate the  mid-infrared (MIR) spectra of many interstellar objects, such as H{\sc ii} regions, planetary, and reflection nebulae, the diffuse ISM, and even entire galaxies. \citep{LegerPuget1984,Allamandola1985,Allamandola1989,Tielens2008,JoblinTielens2011,Li2012}. 
The large heat capacity associated with aromaticity \citep{Schutte1993} suggests that once formed in stellar ejecta \citep{FrenklachFeigelson1989,Cherchneff1992} the PAH molecule is stable against photodissociation, at least compared to other ISM molecules. However, studies of meteorites \citep{Kerridge1987} and recent laboratory experiments \citep{Sandford2000,Thrower2012,Klaerke2013} show that PAHs can undergo processing, such as oxidation, reduction, and isotope exchange, which depend on the environments the PAH experience.
The large size (> 50 C atoms) and numerous hydrogen atoms of interstellar PAHs allows them to potentially be a large reservoir of deuterium in the ISM \citep{Allamandola1989,Tielens1992,Hudgins2004}. Consequently, deuterium-enriched PAHs have been suggested as a possible explanation for the variation of the gas phase atomic D/H ratio in the intermediate region of the Milky Way galaxy, which has an average value of $\sim$ 14 ppm, but has a range of a factor of 4-5  with measurements as low as about 5.0 ppm ($\theta$ Car) and as high as about 21.8 ppm ($\gamma^2$ Vel); the 17 ppm depletion in some regions cannot be explained through astration alone \citep{Peeters2004,Draine2004,Linsky2006,Onaka2014}. \\
\indent  Deuterium atoms can replace hydrogen atom in PAHs and can participate in the same characteristic vibrational modes \citep{Bauschlicher1997}. The heavier mass of deuterium shifts the C-D vibrational modes to longer wavelengths relative to the corresponding C-H vibrational modes \citep{Bauschlicher1997,Allamandola1993, Sandford2000,Hudgins2004}. Hydrogen or deuterium when bonded to the carbon skeleton such that the carbon retains its sp$^2$ hybridization (denoted PAH or PAD, respectively) results in the aromatic C-H stretch mode at 3.29 $\mu$m \citep{Allamandola1989,Sandford2013} or the aromatic C-D stretch mode at 4.40 $\mu$m \citep{Bauschlicher1997,Hudgins2004}. If the hydrogen or deuterium adds to the PAH, resulting in a carbon with sp$^3$ hybridization (denoted H$_n$-PAH or D$_n$-PAH, respectively), then the CH$_2$ or CDH groups show both asymmetric and symmetric aliphatic stretching modes. These features fall at 3.40 and 3.50 $\mu$m for the C-H asymmetric and symmetric stretching modes, respectively, \citep{Sandford2013}, and 4.63 and 4.75 $\mu$m for the C-D asymmetric and symmetric stretching modes \citep{Hudgins2004}. Recently, \citet{Buragohain2015} showed that the 4.75 $\mu$m feature may also be due to the C-D stretch of deuteronated PAH cations (D$_n$-PAH$^+$). For both C-H and C-D aliphatic stretching modes, the intensity of the asymmetric stretch is intrinsically greater than the symmetric stretch because of the larger dipole moment. Other infrared features indicative of deuterated PAHs can arise in the 9-18 $\mu$m as a result of bending of the C-D bonds. However, their exact position will vary across that region depending on the number of nonadjacent D atoms \citep{Peeters2004}, in some cases superimposing on the C-H bending modes of standard PAHs. \\
\indent Detection of PAD/D$_n$-PAHs features in the IR spectrum present the astronomer with several challenges.  All three of the deuterated C-H features are predicted to be weak \citep{Bauschlicher1997,Hudgins2004}. Also, their expected positions coincide with HI emission lines, and the symmetric stretching feature at 4.75 $\mu$m is the only feature to lie in a portion of the spectrum clear of other emission lines. Additionally, ground-based observations of deuterated PAHs are almost impeded by the absorption band of telluric CO$_2$ between 4.3-4.7 $\mu$m \citep{Bauschlicher1997,Hudgins2004}. The best targets for the search of deuterated PAHs are regions where the PAH emission is strongest, such as the surface layers of photodissociation regions (PDRs) in H{\sc ii} regions. \\
\indent  So far, deuterated PAHs have been detected  by ISO-SWS in the Orion Bar and M17 primarily through the C-D asymmetric stretching feature. The estimated number of deuterium atoms relative to hydrogen atoms on PAHs (denoted PAH D/H), based on the ratio of sum of the 4.4 and 4.63 $\mu$m intensities to the 3.29 and 3.4-3.5 $\mu$m intensities, were found to be 0.17 $\pm$ 0.03 in the Orion Bar, and 0.36 $\pm$ 0.08 in M17 \citep{Peeters2004} without considering the intrinsic intensities between the C-D and C-H stretching features, a factor of $\sim$ 1.75 \citep{Bauschlicher1997}. The PAH D/H ratio, based on deuterated PAH features in M17, was indeed consistent with the observed Galactic variation of atomic D/H ratio in the gas phase. Recently, \citet{Onaka2014} reported an upper limit PAH D/H ratio of 0.029 $\pm$ 0.002 in the Orion Bar and 0.023 $\pm$ 0.004 in M17 with AKARI observations at slightly different pointing positions compared to the ISO-SWS observations, and taking the intrinsic intensities into account. The significant difference in the observed deuterium abundance in PAHs made it desirable to obtain multiple spectra of a large number of sources, at a variety of galactic evolution stages, with high signal-to-noise ratios to better constrain the deuterium abundance in PAHs, and to determine if the deuterium fractionation of PAHs could be not only highly variable between sources, but also within a source. To this end, we present our search for deuterated PAHs in a sample of H{\sc ii} regions observed with AKARI. \\
\indent This article is organized as follows. Section 2 contains the details of the NIR spectroscopic observations of 53 H{\sc ii} regions using the AKARI satellite and necessary data reduction method. Section 3 discusses the spectral results, including the first detection of the deuterated PAHs in six Galactic sources, and in Section 4 and 5 the astrophysical implications and conclusion are presented. \\
\section{Observations and data reduction}
\indent The infrared camera (IRC) of the AKARI satellite offers NIR spectroscopy of the 2.5-5 $\mu$m region with a resolution of R $\sim$ 100 for diffuse sources \citep{Onaka2007}. The observations in this study were taken during the post-helium mission phase (Phase 3) of the AKARI satellite with the Nh slit (1'x3") with grism disperser, providing a dispersion of 0.0097 $\mu$m/px in this range \citep{Onaka2007}. At 5 $\mu$m the grism sensitivity decreases compared to 3 $\mu$m region, resulting in a larger noise level in the region where the PAD features are expected. \\
\indent This study is based on the DABUN observational program, which observed seven H{\sc ii} regions  in the Large Magellanic Cloud (LMC), five in the Small Magellanic Cloud (SMC), and eight in the Milky Way (MW), chosen based on their corresponding radio data (\citealt{Paladini2003,Filipovic1998}, respectively). Thirty-three additional Milky Way sources were added from the AKARI Near-Infrared Spectral Atlas of Galactic H{\sc ii} Regions Version 1 public release \citep{Mori2014}. The details of the observation data are given in Table 1. \\
\indent The data reduction was carried out with the official AKARI spectroscopy pipeline for the Phase 3 data (version 20111121; \citealt{Onaka2009}). Spectra were extracted from the area corresponding to the brightest PAH flux at 3.3 $\mu$m. This corresponds to extraction areas of 10''.22 x 3'' for LMC and SMC sources, 7''.3 x 3''  or 10''.22 x 3'' for MW sources that were part of the DABUN program and 8".76 x 3" for MW sources retrieved from the AKARI public release catalog of \citet{Mori2014}. The resulting spectra were subsequently spatially and spectrally smoothed by 3 pixels to remove shot noise without significantly changing the spatial or spectral resolution. In later observations, the thermal noise was noticeable even after pipeline processing, requiring additional post-pipeline dark current subtraction, which was performed following the procedure outlined by \citet{Mori2014}. \\
\indent For most of the targets, we took two or three observations (Table 1). The pointing accuracy of AKARI can vary up to about 30" between the intended and actual target pointing positions; as such we analyzed each observation separately. In the cases with three observations of the same source, the spectrum with significantly different features was removed from analysis based on the assumption that, because of limited pointing accuracy, that spectrum was observing a significantly different part of the H{\sc ii} region. The selection was further reduced by removing the spectra without sufficient signal-to-noise to quantify extinction or PAH emission features. \\
\begin{table*}
\small
\caption{Observation log and target parameters}             
\label{table:1}      
\centering          
\begin{tabular}{l c c l l}     
\hline\hline       
Target & \multicolumn{2}{l}{Slit Center Position$^a$} & Obs.ID & Obs.Date \\ 
          & RA & Dec &       &  \\
\hline                    
    LMCN4A & 73.029 & -66.921 & 4300021.1, 4300021.2, 4301021.1 & 2008 Dec 4, 2008 Dec 4, 2009 Dec 4 \\
    LMCN83B & 73.609 & -69.184 & 4300022.1, 4300022.2, 4301022.1 & 2008 Nov 14, 2008 Nov 14, 2009 Nov 18 \\
    LMCN57A & 83.104 & -67.698 & 4300023.1, 4300023.2, 4301023.2 & 2008 Nov 8, 2008 Nov 8, 2009 Nov 8 \\
    LMCN105A-IRS1 & 77.453 & -68.879 & 4300024.1, 4300024.2, 4301024.2 & 2008 Nov 11, 2008 Nov 11, 2009 Nov 12 \\
    LMCN91A & 74.313 & -68.442 & 4300025.1, 4300025.2, 4301025.3 & 2008 Nov 19, 2008 Nov 20, 2009 Nov 24 \\
    LMCN77A & 72.363 & -69.202 & 4300026.1, 4300026.2, 4301026.3 & 2008 Nov 16, 2008 Nov 16, 2009 Nov 18 \\
    LMCN191A & 76.157 & -70.908 & 4300027.1, 4300027.2, 4301027.4 & 2008 Oct 28, 2008 Oct 28, 2009 Nov 6 \\
    SMCN26 & 12.036 & -73.249 & 4300028.1, 4300028.2 & 2008 Nov 1, 2008 Nov 2 \\
    SMCN10 & 11.235 & -73.170 & 4300029.1, 4300029.2 & 2008 Oct 31, 2008 Nov 1 \\
    SMCN88A & 21.033 & -73.151 & 4300030.1, 4300030.2 & 2008 Nov 4, 2008 Nov 4 \\
    SMCN66 & 14.772 & -72.177 & 4300031.1, 4300031.2, 4301031.1 & 2008 Nov 4, 2008 Nov 5, 2009 Nov 6 \\
    SMCN81 & 17.304 & -73.194 & 4300032.1, 4300032.2 & 2008 Nov 3, 2008 Nov 3 \\
    IRAS14567-5846 &  225.230 & -58.981 & 4300033.1, 4300033.2  & 2009 Feb 19, 2009, Feb 19 \\
    IRAS15384-5348 & 235.569 & -53.976 & 4300034.1, 4300034.2 & 2009 Feb 24, 2009 Feb 24 \\
    IRAS15502-5302 & 238.527 & -53.194 & 4300035.1, 4300035.2 & 2009 Feb 26, 2009 Feb 26 \\
    IRAS12073-6233 & 182.494 & -62.832 & 4300036.1, 4300036.2 & 2009 Jan 28, 2009 Jan 28 \\
    GAL314.2+00.3 & 216.237 & -60.511 & 4300037.1, 4300037.2 & 2009 Feb 14, 2009 Aug 19 \\
    GAL319.9+00.8 & 225.905 & -57.650 & 4300038.1, 4300038.2 & 2009 Feb 19, 2009, Feb 19 \\
    GAL336.0+00.1 & 247.744 & -48.164 & 4300039.1, 4300039.2 & 2009 Mar 3, 2009 Mar 3 \\
    GAL334.7-00.7 & 247.269 & -49.656 & 4300040.1, 4300040.2 & 2009 Mar 3, 2009 Mar 4 \\
    M8    & 270.922 & -24.377 & \multicolumn{1}{l}{5200161.1} & \multicolumn{1}{l}{2008 Sep     23} \\
    G8.137+0.228 & 270.759 & -21.800 & \multicolumn{1}{l}{5200163.1} & \multicolumn{1}{l}{2008  Sep     22} \\
    W31a  & 272.363 & -20.322 & \multicolumn{1}{l}{5200165.1} & \multicolumn{1}{l}{2008 Sep     24} \\
    W31b  & 272.255 & -20.084 & \multicolumn{1}{l}{5200167.1} & \multicolumn{1}{l}{2008 Sep     24} \\
    M17b  & 275.119 & -16.204 & \multicolumn{1}{l}{5200171.1} & \multicolumn{1}{l}{2008 Sep     28} \\
    M17a  &  275.110 & -16.181 & \multicolumn{1}{l}{5200169.1} & \multicolumn{1}{l}{2008        Sep     27} \\
    W42   & 279.564 & -6.795 & 5200294.1, 5200294.2 & \multicolumn{1}{l}{2008   Oct     2, 2008 Oct 2 } \\
    G29.944-0.042 & 281.518 & -2.653 & 5200295.1, 5200295.2 & \multicolumn{1}{l}{2008   Oct     4, 2008 Oct 4} \\
    W49A  & 287.568 & 9.108 & 5200299.1, 5200299.2 & \multicolumn{1}{l}{2008    Oct     12, 2007 Oct 13} \\
    G48.596+0.042 & 290.127 &  13.930 & 5200300.1, 5200300.2 & \multicolumn{1}{l}{2008  Oct     17, 2008 Oct 17} \\
    W51   & 290.561 & 14.051 & 5200301.1, 5200301.2 & \multicolumn{1}{l}{2008   Oct     17, 2008 oct 18} \\
    W58A  &  300.440 & 33.548 & 5201198.1, 5200767.1 & \multicolumn{1}{l}{2009  May     2, 2009 Nov 6} \\
    G70.293+1.600 & 300.440 & 33.548 & \multicolumn{1}{l}{5200337.1} & \multicolumn{1}{l}{2008  Nov     6} \\
    G75.783+0.343 & 305.422 &  37.430 & \multicolumn{1}{l}{5200772.1} & \multicolumn{1}{l}{2009 May     11} \\
    G76.383-0.621 & 306.863 & 37.381 & \multicolumn{1}{l}{5200343.1} & \multicolumn{1}{l}{2008  Nov     15} \\
    G78.438+2.659 & 304.913 & 40.943 & \multicolumn{1}{l}{5200776.1} & \multicolumn{1}{l}{2009  May     13} \\
    DR7   & 307.037 & 40.875 & \multicolumn{1}{l}{5200769.1} & \multicolumn{1}{l}{2009  May     17} \\
    G81.679+0.537 & 309.752 & 42.331 & \multicolumn{1}{l}{5200347.1} & \multicolumn{1}{l}{2008  Nov     22} \\
    G111.282-0.663 &  349.020 & 60.038 & \multicolumn{1}{l}{5200432.1} & \multicolumn{1}{l}{2009 Jan     16} \\
    RCW42 & 141.106 & -51.990 & \multicolumn{1}{l}{5200452.1} & \multicolumn{1}{l}{2008 Dec     15} \\
    G282.023-1.180 & 151.653 & -57.204 & \multicolumn{1}{l}{5200436.1} & \multicolumn{1}{l}{2009 Jan     1} \\
    RCW49 & 156.034 & -57.788 & \multicolumn{1}{l}{5200438.1} & \multicolumn{1}{l}{2009 Jan     4} \\
    NGC3372 & 160.883 & -59.580 & \multicolumn{1}{l}{5200440.1} & \multicolumn{1}{l}{2009       Jan     10} \\
    G289.066-0.357 & 164.124 & -60.098 & \multicolumn{1}{l}{5200442.1} & \multicolumn{1}{l}{2009 Jan     13} \\
    NGC3576 & 167.984 & -61.313 & \multicolumn{1}{l}{5200444.1} & \multicolumn{1}{l}{2009       Jan     17} \\
    NGC3603 & 168.756 & -61.263 & \multicolumn{1}{l}{5200446.1} & \multicolumn{1}{l}{2009       Jan     17} \\
    G319.158-0.398 & 225.816 & -59.074 & \multicolumn{1}{l}{5200933.1} & \multicolumn{1}{l}{2009 Aug     25} \\
    G330.868-0.365 & 242.601 & -52.099 & \multicolumn{1}{l}{5200109.1} & \multicolumn{1}{l}{2008 Sep     2} \\
    G331.386-0.359 & 243.183 & -51.748 & \multicolumn{1}{l}{5200113.1} & \multicolumn{1}{l}{2008 Sep     3} \\
    G333.122-0.446 & 245.255 & -50.585 & \multicolumn{1}{l}{5200121.1} & \multicolumn{1}{l}{2008 Sep     4} \\
    G338.398+0.164 & 250.032 & -46.385 & 5200942.1, 5200942.2 & \multicolumn{1}{l}{2009 Sep     7, 2009 Sep 7} \\
    G338.400-0.201 & 250.468 & -46.582 & \multicolumn{1}{l}{5200943.2} & \multicolumn{1}{l}{2009 Sep     7} \\
    G345.528-0.051 & 256.538 & -40.962 & \multicolumn{1}{l}{5200133.1} & \multicolumn{1}{l}{2008 Sep     11} \\
\hline 
     \multicolumn{4}{l}{a. Intended AKARI target position in degrees, J2000} &\\                
\end{tabular}
\end{table*}
\section{Results and analysis}
\indent All of the analyzed spectra show  a number of features, typical of H{\sc ii} regions, such as HI recombination lines, CO$_2$ ice features, and the PAH bands at 3.29 and 3.4-3.6 $\mu$m (Figure 1). The uncertainty in the relative flux calibration is less than 10\%.  The MW sources have better signal-to-noise compared to the LMC and SMC sources, which lie on average at distances of $\sim$ 50 kpc \citep{Pietrzynski2013} and $\sim$ 60 kpc \citep{Hilditch2005}, respectively. As a result, while the PAH aromatic C-H stretching mode at 3.29 $\mu$m is seen in all of the spectra, in some of the LMC and SMC spectra the PAH flux is too weak to distinguish the aliphatic C-H features from the noise. \\
\indent Some of the MW source spectra show ice absorption features of H$_2$O and CO$_2$, which likely arise from cold interstellar clouds between the source and AKARI. All of the obtained spectra show emission features indicative of ionized gas in H{\sc ii} regions. For example, the prominent HI recombination lines Br$\alpha$ at 4.052 $\mu$m and Br$\beta$ at 2.626 $\mu$m are seen in all spectra, and a few also show a number of other hydrogen and helium recombination lines, all of which are fit with Gaussian functions;  the fit parameters are listed in Table 2. There is a shift in the observed central wavelength of the HI lines relative to their literature values, but the discrepancy is within the uncertainty of the wavelength calibration of $\sim$ 0.005 $\mu$m. \\
\indent The continuum is fit with a 3rd order polynomial, taking into account the broad continuum plateau from 3.2-3.6 $\mu$m, and then subtracted. The H$_2$O absorption feature around 3.05 $\mu$m is fit via laboratory spectrum of pure H$_2$O ice at 10 K taken from the Leiden Ice Database \citep{Gerakines1996}. In contrast, the CO$_2$ ice feature cannot be completely resolved using the AKARI/IRC slit spectroscopy, and is fit using a negative Gaussian function; details of the fit for the spectra are listed in Table 2. \\
\begin{table}[h]
\caption{Gaussian profile parameters for ice absorption features and emission lines fitted in the spectra}             
\label{table:2}      
\centering                          
\begin{tabular}{l c c}        
\hline\hline                 
Line & $\lambda_{center}$ ($\mu$m) & FWHM ($\mu$m) \\    
\hline                        
HI Br$\beta$ & 2.6259 & 0.026 \\
HI Pf13 & 2.6751 & 0.026 \\
HI Pf12 & 2.7583 & 0.026 \\
HI Pf$\eta$ & 2.8730 & 0.026 \\
HI Pf$\epsilon$ & 3.0392 & 0.026 \\
PAH & 3.29 & 0.060 \\
HI Pf$\delta$ & 3.2970 & 0.026 \\
H$_n$-PAH & 3.40 & 0.058 \\
H$_n$-PAH & 3.45 & 0.058 \\
H$_n$-PAH & 3.50 & 0.058 \\
H$_n$-PAH & 3.56 & 0.058 \\
HI Pf$\gamma$ & 3.7406 & 0.026 \\
H$_2$ 0-0 S(13) & 3.846 & 0.026 \\
HI Br$\alpha$ & 4.0523 & 0.026 \\
HI Hu13 & 4.1708 & 0.026 \\
CO$_2$ & 4.26 & 0.047 \\
He I ($^3$S$_1$-$^3$P$_0$) & 4.2954 & 0.026 \\
HI Hu12 & 4.3765 & 0.026 \\
PAD & 4.40 & 0.047 \\
D$_n$-PAH & 4.63 & 0.047 \\
HI Pf$\beta$ & 4.6538 & 0.026 \\
HI Hu$\epsilon$ & 4.6725 & 0.026 \\
D$_n$-PAH & 4.75 & 0.047 \\
D$_n$-PAH & 4.80 & 0.047 \\
D$_n$-PAH & 4.85 & 0.047 \\
\hline                                   
\end{tabular}
\end{table}
\begin{figure*}
  \centering
            \includegraphics[width=18cm]{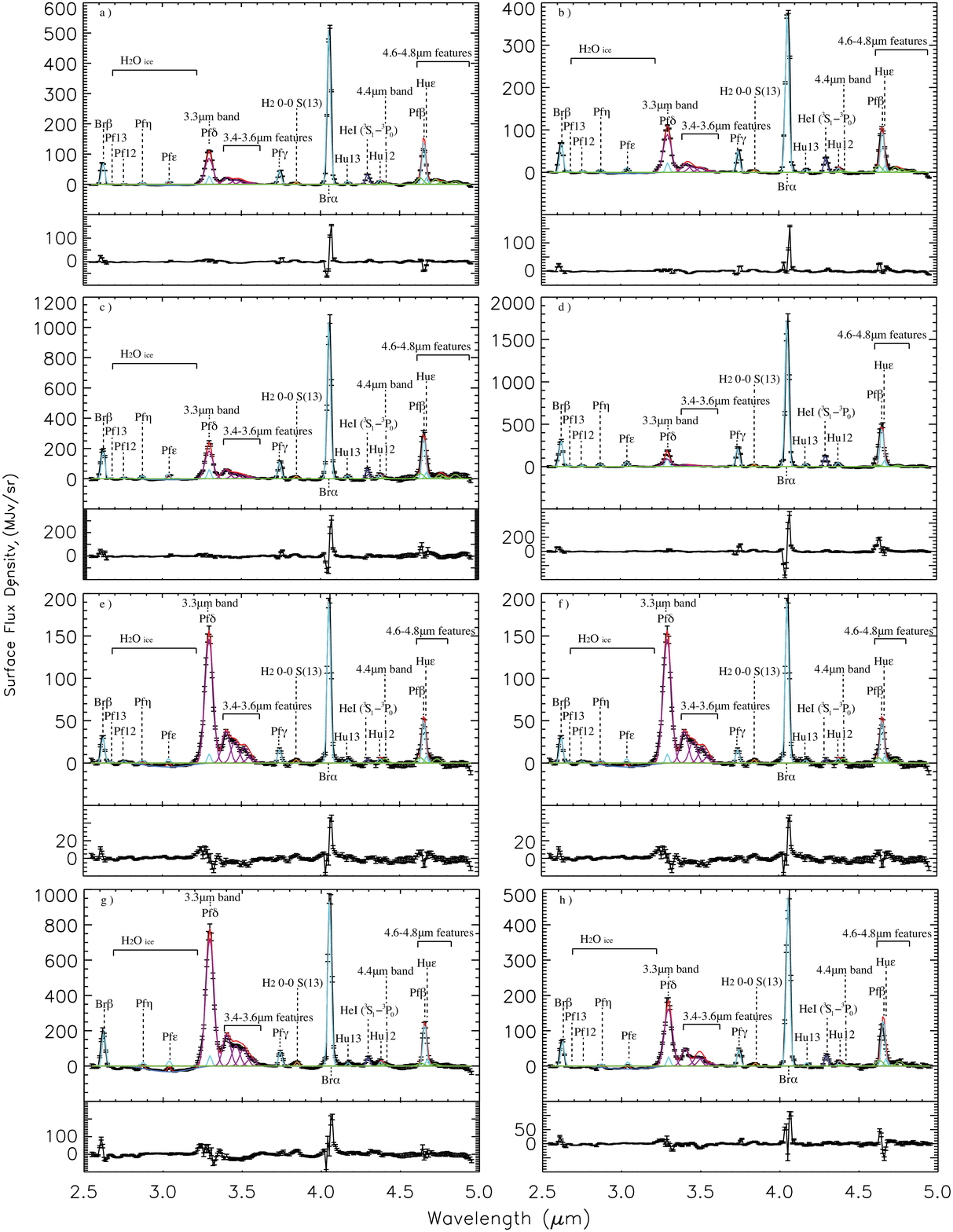}
      \caption{Fitting results for the spectra of a) IRAS12073-6233 obs. 1, b) IRAS12073-6233 obs. 2, c) NGC3603, d) M17b, e) W51 obs. 1, f) W51 obs. 2, g) M8, and h) G75.783+0.343 in red. The black line is the observed spectra, the HI emission lines are fit in cyan, the HeI emission line is fit in navy, the H$_2$O ice absorption line is fit in blue, the H$_2$ rotational line is fit in orange, PAH and H$_n$-PAH features are fit in purple, and PAD and D$_n$-PAH features are fit in green. Below each figure is the corresponding residual plot.
              }
         \label{Fig1}
\end{figure*}
\begin{figure*}
  \centering
            \includegraphics[width=18cm]{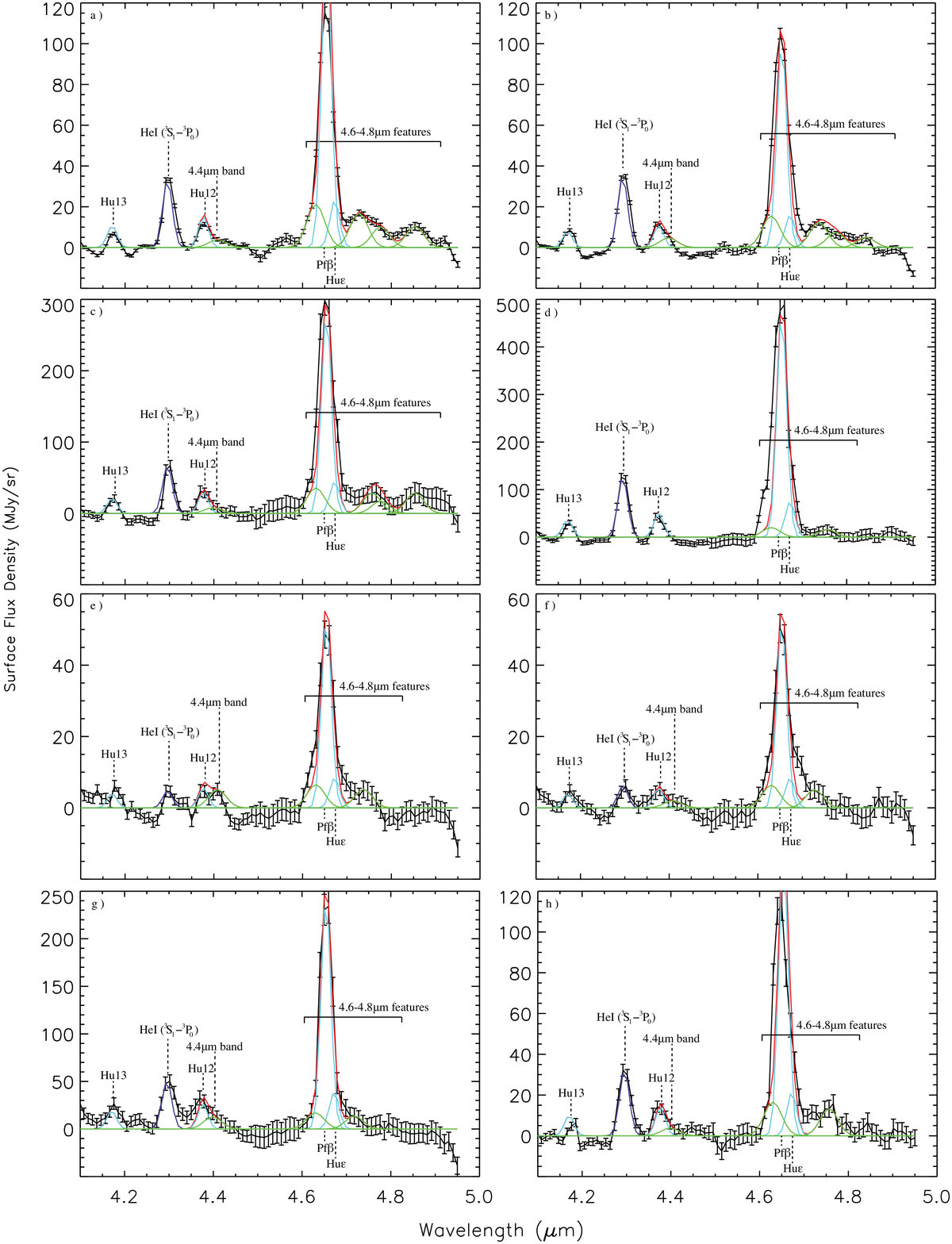}
      \caption{Close up of the C-D stretching region of the spectra of a) IRAS12073-6233 obs. 1, b) IRAS12073-6233 obs. 2, c) NGC3603, d) M17b, e) W51 obs. 1, f) W51 obs. 2, g) M8, and h) G75.783+0.343. The fitting results for the spectra is shown in red, the black line is the observed spectra, the HI emission lines are fit in cyan, the HeI emission line is fit in navy, and PAD and D$_n$-PAH features are fit in green.
              }
         \label{Fig2}
\end{figure*}
\indent The PAH and D$_n$-PAH fluxes at 3.29 and 4.63 $\mu$m, respectively, overlap with the HI emission lines Pf$\delta$ at 3.297 $\mu$m, Pf$\beta$ at 4.654 $\mu$m, and Hu$\epsilon$ at 4.673 $\mu$m.  The expected HI emission intensities are determined relative to Br$\alpha$ and Br$\beta$, assuming case B recombination conditions of T$_e$ = 10$^4$ K and n$_e$ = 10$^4$ cm$^{-3}$ for the Milky Way and T$_e$ = 10$^4$ K and n$_e$ = 10$^2$ cm$^{-3}$ for the LMC and SMC \citep{StoreyHummer1995} and an extinction law of A$_{\lambda}$ = $\lambda^{-1.7}$ \citep{MartinHernandez2002a}. To determine the flux of the underlying features, the overlapping HI lines are modeled as Gaussian functions and subtracted from the spectrum. The extinction-corrected intensity subtraction introduces an uncertainty of $\sim$ 10\% to the flux of the underlying feature. The observed flux at 4.17 and 4.37 $\mu$m can be fit by the HI Hu13 and Hu12 expected flux, with some excess flux at 4.4 $\mu$m. \\
\indent The PAH and H$_n$-PAH features and their deuterium counterparts are fit with Gaussian functions; the details of the fits are given in Table 2. Based on band position coincidence between observed excess flux and laboratory \citep{Sandford2000} and theoretical data \citep{Hudgins2004}, PAD and/or D$_n$-PAH features were detected in six Milky Way sources: IRAS12073-6233, NGC3603, M8, M17b, W51, and G75.783+0.343 (Figures 1 and 2); the calculated PAH and H$_n$-PAH fluxes are summarized in Table 3, while the calculated PAD and D$_n$-PAH fluxes are summarized in Table 4. Of the six sources, IRAS12073-6233 and W51 had two observations that showed PAD and D$_n$-PAH features. In addition three sources: IRAS 15384-5348, M17a, and NGC3576, show a less than 3 $\sigma$ detection of the asymmetric C-D stretch, but because of noisy baseline the accompanying symmetric mode was not seen. Aromatic C-H stretch overtone and combination bands, as well as, contributions from aliphatic side groups also fall in the range of 3.4-3.6 $\mu$m \citep{Allamandola1989, Sandford2013,Pilleri2015}, and similarly, contributions from the C-D analogs likely fall between 4.6-5.0 $\mu$m. For analysis, we assume that aliphatic groups are attached to the main PAH rings, \textit{i.e.,} superhydrogenated PAHs/PADs contribute the majority of the flux in these wavelength ranges, and the determined strength of these bands are consequently upper limits. \\
\indent The aromatic C-D stretch feature at 4.4 $\mu$m and aliphatic C-D (a)symmetric stretch features at 4.63-4.85 $\mu$m are present in all eight of the spectra (Figures 1 and 2), with the exception of M17b. As noted by \citet{Peeters2004} for M17, the nearby source M17b does not show the aromatic C-D stretch band. The aromatic C-D stretch mode is observed to be blended with the HI Hu12 emission line. Nonetheless, two Gaussian profiles are needed to reproduce the observed spectrum, and detections of the same blended feature in multiple sources suggest that the feature is not noise even though they are only  $\sim$ 1 $\sigma$ level detections. The aliphatic C-D asymmetric stretch feature is heavily blended with HI emission lines and, as a result, its intensity is an estimation based on the intrinsic intensity ratio of the C-D symmetric to asymmetric modes and the intensity of the unobscured C-D symmetric feature. Simultaneous fitting of the Pf$\beta$, Hu$\epsilon$, and estimated C-D asymmetric stretch features is able to reproduce the observed spectrum, which suggests that the estimation is good. In the case of IRAS12073-6233 and NGC3603 (Figure 1 a), b), c) and Figure 2 a), b), c)), the PAD and D$_n$-PAH signal to noise was large enough that the minor features seen in the \citet{Hudgins2004} modeled spectra at 4.8-4.9 $\mu$m are discernible.  In NGC3603, however, the minor feature at 4.84 $\mu$m has a significantly larger intensity than expected based on the model in \citet{Hudgins2004}, which is likely a result of the continuum subtraction. \\
\indent In the 3.8-4.6 $\mu$m region, there are a number of features that can be attributed to either HI lines, the 0-0 S(13)-S(9) ro-vibrational transitions of H$_2$, or deuterated PAHs. Notably, the S(10) transition of H$_2$ at 4.4 $\mu$m overlaps the expected position of the aromatic C-D stretch mode. The S(13) transition at 3.85 $\mu$m is seen clearly in all eight of the spectra. Based on the intensities of the excess flux at the positions corresponding to the S(12) - S(9) transitions  of H$_2$ and level populations predicted by non-LTE models of H$_2$ \citep{Bertoldi2000}, we cannot rule out the S(10) transition of H$_2$ as the carrier of the 4.4 $\mu$m excess flux at the present spectral resolution. However, for the analysis we assume the 4.4 $\mu$m feature is the aromatic C-D stretch in order to obtain an upper limit on its abundance. In the case of the C-D aliphatic stretch features, the excess fluxes at 4.63 $\mu$m, and 4.75 $\mu$m do not match the S(9) transition of H$_2$ within the wavelength calibration uncertainties, and thus the H$_2$ transition does not influence their assignments. \\ 
\indent If we only take the intrinsic intensity ratio of C-D to C-H features into account, which reduces the overall intensities for deuterium features by a factor of $\sim$ 1.75 \citep{Bauschlicher1997}, the number of deuterium atoms relative to hydrogen atoms on PAHs is then estimated from the ratio of the sum of the corrected deuterated features to the sum of the hydrogenated features. An observed upper limit of the PAH D/H is determined to be between 0.03 and 0.44, which is summarized in Table 4. \footnote{The given PAH D/H uncertainties in Table 4 do not take  errors of various origins into account, including the difference of excitation of PADs/D$_n$-PAHs and PAHs/H$_n$-PAHs, the assignment uncertainties of minor spectral features, or the uncertainties in the intrinsic intensities between different vibrational modes.} For sources with multiple observations, W51 and IRAS12073-6233, it was found that the PAH D/H ratio is consistent within flux uncertainties in W51, but not in IRAS12073-6233. This may hint that very local conditions are indeed important to the deuterium variations, but more observations at higher spatial resolution are needed to confirm this. \\
\indent The amount of deuterium at either an aromatic or aliphatic position was also determined through a comparison of observed PAH D/H ratios to those expected if one hydrogen is replaced with a deuterium at either an aromatic or aliphatic position. From the 3.4/3.29 $\mu$m ratio, the amount of aliphatic hydrogen relative to aromatic hydrogen, N$_{aliphatic,H}$/N$_{aromatic,H}$, of the PAHs in the eight observations was determined to be between 0.2 - 0.3 (Table 3), which is significantly larger than the N$_{aliphatic,H}$/N$_{aromatic,H}$ $\sim$ 0.02 typically seen in PAHs in the ISM \citep{Tielens2008}. Additionally, neutral, small (N$_C$ < 50) PAHs molecules are considered to mainly contribute to the 3.3 $\mu$m feature, so from the PAH IR Spectral Database \citep{PAHDB} six representative PAHs (three extended and three compact) were chosen to determine the expected D/H ratios: C$_{40}$H$_{22}$, C$_{40}$H$_{16}$, C$_{32}$H$_{18}$, C$_{32}$H$_{14}$, C$_{24}$H$_{14}$, and C$_{24}$H$_{12}$. For example, for C$_{40}$H$_{22}$ to get an N$_{aliphatic,H}$/N$_{aromatic,H}$ $\sim$ 0.22, there would be four aliphatic hydrogens and 18 aromatic hydrogens. Replacement of an aromatic hydrogen with an aromatic deuterium would result in an aromatic D/H of 0.06, and replacement of one aliphatic hydrogen for an aliphatic deuterium would result in an aliphatic D/H ratio of 0.33. If this is evaluated for the six representative molecules, on average a monodeuterated D$_n$-PAH would have an expected aliphatic D/H of $\sim$ 0.5, while a monodeuteratued PAD would have an average expected aromatic D/H of $\sim$ 0.09. \\
\indent The ratio of the 4.40 $\mu$m features to the 3.29 $\mu$m features gives a fractional abundance of aromatic deuterium to aromatic hydrogen of < 0.05 for all eight observations. In contrast, based on the sum of the 4.63 and 4.75 $\mu$m features relative to the sum of the 3.40 and 3.50 $\mu$m features, the fractional abundance of aliphatic deuterium relative to aliphatic hydrogen is much larger, varying from 0.09 to 1. For a representative PAH, for example, the hydrogenated PAH C$_{40}$H$_{18}$, these observed deuterium abundances would yield at most 1 aromatic deuterium, 2 aliphatic deuterium, 13 aromatic hydrogen, and 2 aliphatic hydrogen, and a PAH D/H of 0.2, which is roughly representative of the average of the values determined for the eight observations. Based on the expected D/H ratios, it is estimated that in all eight observations less than 10\% of the observed PAHs have one aromatic deuterium. Conversely, with the exception of the PAHs in W51 and M8, all of the observed PAHs have at least one aliphatic deuterium. In particular, the aliphatic D/H ratios for IRAS12073-6233 and M17b are more in agreement with the PAHs having one aliphatic deuterium for every aliphatic hydrogen. Furthermore, the aliphatic D/H ratios $\sim$ 0.4 and $\sim$ 0.30 for IRAS12073-6233 obs. 1 and obs. 2, respectively, suggest that almost all of the deuterium is in aliphatic positions. \\
\indent The determined amount of deuterium relative to hydrogen on PAHs are upper limits, and a more complete understanding would include the effects of the emission process on the band intensities; to understand these effects we calculated the emission spectrum of a prototype PAH molecule, neutral ovalene, where one solo hydrogen is substituted with a -CH$_2$D group (N$_C$ = 33). To model the emission process, we follow the procedure described in \citet{Candian2012}, where both the infrared spectrum of the molecule in question and its photoabsorption cross-section were evaluated with quantum chemistry techniques. As the effective temperature of the exciting source, we assume T$_{eff}$ = 40000 K, as in IRAS 12073-6233 \citep{MartinHernandez2002b}, which is one of the H{\sc ii} regions in our sample. \\
\indent For this molecule, the intrinsic  3.3/4.75  $\mu$m (C-H aromatic/C-D aliphatic) ratio is 1.81 (consistent with earlier calculations of \citealt{Bauschlicher1997}). The emission process brings the band ratio to 0.64, which then will correspond to a D/H range of 0.01-0.16 in our sample. These values are clearly sensitive to the parameters assumed in the emission model (e.g., effective temperature, PAH size), for example, M17. \citet{Peeters2004} did not consider the emission process and found a D/H = 0.36 $\pm$ 0.08, while for the same source \citet{Onaka2014} via an emission model that favored larger PAHs found an upper limit D/H = 0.023 $\pm$ 0.004. Similarly, we find a D/H = 0.09 $\pm$ 0.05 for the nearby source M17b via our emission model, which favors smaller PAHs. Therefore, the observed band intensity ratio can overestimate the actual relative abundance of deuterium to hydrogen on PAHs and the amount is dependent on the emission process. Stronger constraints on the typical PAH size population and exciting source characteristics in H{\sc ii} regions would improve our estimate of the deuterium abundance. \\

\begin{table*}[h]
\caption{PAH/H$_n$-PAH Fluxes$^a$ for sources with detectable deuterated features}             
\label{table:3}      
\centering          
\begin{tabular}{l c l l l l l }     
\hline\hline       
Source           &  Av (mag)      &  3.29$\mu$m &  3.4$\mu$m & 3.45$\mu$m$^b$ & 3.5$\mu$m & 3.56$\mu$m$^b$ \\
                         &                        &  Aromatic CH& Asymmetric & & Symmetric & \\
                         &                        &                      & Aliphatic CH & & Aliphatic CH & \\
                         
\hline
    G75.783+0.343        &  28   &  22.2 $\pm$ 2.0  &  6.2 $\pm$ 0.6  &  3.3 $\pm$ 0.4  &  2.9 $\pm$ 0.9  &  0.9 $\pm$ 0.2  \\
                    &        &  125 $\sigma$  &  37 $\sigma$  &  20 $\sigma$  &  18 $\sigma$  &  5.5 $\sigma$  \\
    NGC3603              &  22   &  25.1 $\pm$ 2.4  &  7.8 $\pm$ 0.8  &  4.6 $\pm$ 0.6  &  3.1 $\pm$ 1.2  &  1.2 $\pm$ 0.5  \\
                    &        &  69 $\sigma$  &  23 $\sigma$  &  14 $\sigma$  &  9.5 $\sigma$  &  3.7 $\sigma$  \\
    W51 obs. 1           &  26   &  20.0 $\pm$ 1.8  &  4.8 $\pm$ 0.5  &  3.2 $\pm$ 0.3  &  2.3 $\pm$ 0.5  &  0.9 $\pm$ 0.2  \\
                    &        &  182 $\sigma$  &  47 $\sigma$  &  32 $\sigma$  &  23 $\sigma$  &  9.5 $\sigma$  \\
    W51 obs. 2           &  26   &  20.6 $\pm$ 1.8  &  4.9 $\pm$ 0.5  &  3.2 $\pm$ 0.3  &  2.2 $\pm$ 0.4  &  1.2 $\pm$ 0.2  \\
                    &        &  232 $\sigma$  &  59 $\sigma$  &  39 $\sigma$  &  27 $\sigma$  &  15 $\sigma$  \\                 
    M17b                 &  23   &  12.3 $\pm$ 1.5  &  2.9 $\pm$ 0.5  &  2.1 $\pm$ 0.4  &  0.8 $\pm$ 0.5  &  0.8 $\pm$ 0.4  \\
                    &        &  20       $\sigma$  &  5.1 $\sigma$  &  3.6 $\sigma$  &  1.5 $\sigma$  &  1.4 $\sigma$  \\
    M8            &  16  &  99.1 $\pm$ 8.9  &  22.9 $\pm$ 2.2  &  14.7 $\pm$ 1.5  &  12.1 $\pm$ 2.2  &  5.5 $\pm$ 0.9  \\
                    &        &  175 $\sigma$  &  44 $\sigma$  &  28 $\sigma$  &  24 $\sigma$  &  11 $\sigma$  \\
    IRAS12073-6233 obs. 1        &  31   &  11.7 $\pm$ 1.0  &  2.9 $\pm$ 0.3  &  1.9 $\pm$ 0.2  &  1.3 $\pm$ 0.2  &  0.6 $\pm$ 0.1  \\
                    &        &  149 $\sigma$  &  40 $\sigma$  &  27 $\sigma$  &  18 $\sigma$  &  9.0 $\sigma$  \\ 
    IRAS12073-6233 obs. 2        &  22   &  12.2 $\pm$ 1.0  &  2.6 $\pm$ 0.2  &  1.7 $\pm$ 0.2  &  1.3 $\pm$ 0.2  &  0.5 $\pm$ 0.1  \\
                    &        &  69 $\sigma$  &  16 $\sigma$  &  11 $\sigma$  &  8.0 $\sigma$  &  2.8 $\sigma$  \\                  
\hline                  
\multicolumn{2}{l}{a. In units of 10$^{-17}$ Wm$^{-2}$ arcsec$^{-2}$} & & & & &\\
\multicolumn{7}{l}{b. See text for discussion of the origin of these features.} \\  
\end{tabular}
\end{table*}
\begin{table*}[h]
\caption{PAD/D$_n$-PAH Fluxes$^a$ for sources with detectable deuterated features}             
\label{table:4}      
\centering          
\begin{tabular}{l l l l l l l}     
\hline\hline       
  Source                 &  4.4$\mu$m &  4.63$\mu$m &  4.75$\mu$m &  4.8$\mu$m$^b$ & 4.85$\mu$m$^b$ & D/H$^c$\\
                         & Aromatic CD & Asymmetric & Symmetric & & \\
                         &                       & Aliphatic CD & Aliphatic CD & & \\

\hline
    G75.783+0.343        &  0.30 $\pm$ 0.19  &  1.26 $\pm$ 0.53  &  0.99 $\pm$ 0.33  &   & &  0.13 $\pm$ 0.03\\
                    &  1.2 $\sigma$  &  5.3 $\sigma$  &  4.2 $\sigma$  & &    &\\
    NGC3603              &  0.62 $\pm$ 0.52  &  2.73 $\pm$ 1.57  &  2.13 $\pm$ 0.98  &  1.73 $\pm$ 0.79  &  2.11 $\pm$ 1.07  & 0.37 $\pm$ 0.10 \\
                    &  1.3 $\sigma$  &  5.8 $\sigma$  &  4.7 $\sigma$  &  3.8 $\sigma$  &  4.7 $\sigma$  & \\
    W51 obs. 1           &  0.41 $\pm$ 0.14  &  0.49 $\pm$ 0.29  &  0.39 $\pm$ 0.18  &  & & 0.07 $\pm$ 0.02\\
                    &  5.4 $\sigma$  &  6.8 $\sigma$  &  5.5 $\sigma$  &   & & \\
    W51 obs. 2           &  0.16 $\pm$ 0.14  &  0.48 $\pm$ 0.29  &  0.38 $\pm$ 0.18  &  & & 0.06 $\pm$ 0.02\\
                    &  2.1 $\sigma$  &  6.6 $\sigma$  &  5.3 $\sigma$  &   & & \\
    M17b                 &    &  1.53 $\pm$ 1.19  &  1.19 $\pm$ 0.75  &   &  & 0.25 $\pm$ 0.13\\
                    &   &  6.6 $\sigma$  &  5.3 $\sigma$  &   &  &  \\
    M8            &  1.03 $\pm$ 0.53  &  1.05 $\pm$ 1.00  &  0.82 $\pm$ 0.63  &   &  & 0.03 $\pm$ 0.01\\
                    &  4.5 $\sigma$  &  4.8 $\sigma$  &  3.8 $\sigma$  &    &   & \\
    IRAS12073-6233 obs. 1        &  0.30 $\pm$ 0.08  &  1.62 $\pm$ 0.28  &  1.26 $\pm$ 0.17  &  0.76 $\pm$ 0.15  &  0.76 $\pm$ 0.16 & 0.44 $\pm$ 0.05\\
                    &  1.2 $\sigma$  &  6.7 $\sigma$  &  5.3 $\sigma$  &  3.2 $\sigma$  &  3.3 $\sigma$ & \\
    IRAS12073-6233 obs. 2        &  0.37 $\pm$ 0.08  &  1.18 $\pm$ 0.23  &  0.92 $\pm$ 0.23  &  0.46 $\pm$ 0.16  &  0.36 $\pm$ 0.14 & 0.31 $\pm$ 0.04\\
                    &  3.2 $\sigma$  &  10.7 $\sigma$  &  8.5 $\sigma$  &  4.3 $\sigma$  &  3.4 $\sigma$ & \\
\hline            
\multicolumn{2}{l}{a. In units of 10$^{-17}$ Wm$^{-2}$ arcsec$^{-2}$} & & & & &\\
\multicolumn{7}{l}{b. See text for discussion of the origin of these features.} \\  
\multicolumn{7}{l}{c. The PAH D/H is calculated as the sum of the deuterium feature fluxes (Table 4 columns 2-6) divided by a factor } \\ 
\multicolumn{7}{l}{of 1.75 to account for intrinsic intensities, divided by the sum of the hydrogen feature fluxes (Table 3 columns 3-7). } \\  
\multicolumn{7}{l}{The quoted uncertainties are based  on the flux uncertainties alone.} \\   
\end{tabular}
\end{table*}
\section{Discussion}
\indent High levels of deuteration have been observed in some species (e.g., CD$_3$OH, ND$_3$; \citet{Parise2004,Lis2002,vanderTak2002}), which are thought to originate from grain surface chemistry \citep{Roberts2003}. Deuterium fractionation is not as extensive for PAHs. In the best case (IRAS12073-6233 observation 1), the PAH D/H value is 0.44 (Table 4), which would translate to a fraction of gas-phase atomic deuterium (relative to hydrogen) locked up in PAHs of roughly 18 ppm; these are determined following the analysis method outlined in \citet{Onaka2014}. Of the sources with observed deuterated PAHs, the average PAH D/H fraction is 0.21, which corresponds to a locked up fraction of gas-phase deuterium (relative to hydrogen) of about 10 ppm. Observation of atomic deuterium in the local ISM shows strong variation in the D/H abundance ratio of the gas phase at the 17 ppm level \citep{Linsky2006}. While our observations are not along the same lines of sight, they indicate that PADs and D$_n$-PAHs would be a major reservoir of elemental deuterium. Moreover, our study also reveals strong variation in the deuterated PAH to PAH ratio. Hence, the interaction of atomic deuterium with PAHs could well be at the origin of the observed variation in the gas-phase deuterium abundance. \\
\indent While deuterated PAHs are not omnipresent, when present, deuteration is efficient; all eight observations have a PAH D/H ratio that is significantly greater than the cosmic gas-phase abundance of $\sim$ 10$^{-5}$ \citep{VidalMadjar1998}. Incorporation of deuterium into PAHs can occur through a number of mechanisms, most of which are driven by the small difference in zero-point energy between hydrogen and deuterium. Deuterium enrichment can take place in the gas phase or through solid state reactions within ice or on grains. \\
\indent Large deuteration fractionation can occur in PDRs at depths where most of the hydrogen is locked up in H$_2$, but deuterium is still mainly atomic. H$_2$ and HD are expected to show different behavior with depth into a cloud, as self-shielding of H$_2$ pulls the H/H$_2$ transition to the surface of the PDR. Conversely, self-shielding is of little importance for HD, and photodestruction converts HD to atomic deuterium \citep{Tielens1992}. Through gas-phase addition reactions, the free atomic deuterium adds aliphatically to the PAH molecule. Recent theoretical calculations \citep{Ricca2007,Rauls2008,Rasmussen2011} and experiments \citep{Thrower2012,Klaerke2013,Demarais2014} demonstrate that hydrogenation (H$_n$-PAH) or deuteration (D$_n$-PAH) can be important through reactions in the gas phase in regions of the PDR without intense UV radiation. The addition occurs preferentially on carbons at the edges of PAH molecules and gives the carbon an aliphatic character \citep{Rauls2008}.  \citet{Rasmussen2011} and \citet{Rauls2008} calculated  the first hydrogen addition to the periphery has a barrier of 0.06 eV for neutral PAHs, while the second hydrogen addition is barrierless. For cations, hydrogenation is even easier, since the first hydrogen addition is barrierless and the second addition has a negligible barrier \citep{Ricca2007}. Similar mechanisms can be employed to explain the presence of aliphatic deuterium on PAHs. In an evaporating flow, the PAHs move to the surface of the PDR and then into the general ISM, during which the deuterium fractionation is temporarily preserved. In this schematic way, we can understand how PAHs can be highly fractionated,  that this fractionation primarily occurs as aliphatic H/D, and that this fractionation behavior is very sensitive to the local conditions and history of the PAHs. Thus, it is expected to vary from one region to the next. \\
\indent Additionally, ion-molecule and neutral-neutral addition reactions occur at low temperatures, but require that the reaction is exothermic or has no barrier \citep{DalgarnoLepp1984,Tielens1992,Bauschlicher1998,Sandford2001}. These reactions, in the gas phase or on grain surfaces, are proposed to lead to both aliphatic and aromatic deuterated or hydrogenated PAHs deep inside dense clouds, \\
\\
PAH + H$_2$D$^+$ $\longrightarrow$ D$_n$-PAH$^+$ +H$_2$\\
PAH + H$_3^+$ $\longrightarrow$ H$_n$-PAH$^+$ +H$_2$\\
H$_n$-PAD$^+$ + e$^-$ $\longrightarrow$ PAD + H\\
D$_n$-PAH$^+$ + e$^-$ $\longrightarrow$ PAH + D\\
H$_n$-PAH$^+$ + e$^-$ $\longrightarrow$ PAH + H.\\
\\
Similar reaction schemes are responsible for the deuteration of small hydrocarbon species, such as HCO$^+$ and HCN. Deuteration fractionation in these species is observed to reach levels of $\sim$ 4x10$^{-2}$ \citep{Roberts2002}. \\
\indent At temperatures less than 50 K, most of the volatile molecules are frozen out onto the dust grains \citep{Boogert2015}. Penetrating UV radiation from nearby O/B stars or photon-induced, cosmic-ray ionization that is deep inside dense cores has enough energy to break the molecular bonds on smaller molecules producing radicals. These highly reactive species, in turn, can go on to form new bonds \citep{Bernstein2001,Sandford2001}. Laboratory experiments of PAHs in deuterium-enriched water ices demonstrated that under UV irradiations PAHs undergo oxidation, reduction, and deuterium-hydrogen exchange reactions. Deuterium enrichment in ices is independent of the size of the PAH, and seems to favor the aromatic deuterium product (PAD) over aliphatic addition, resulting in enrichment levels of at least D/H $\sim$ 10$^{-4}$ \citep{Sandford2000}. \\
\indent Independent of the temperature, PAHs can undergo unimolecular photodissociation if they absorb a UV photon with enough energy to break the C-H bond. The aliphatic sp$^3$ bonds are more labile compared to the aromatic sp$^2$ bonds, favoring the loss of an aliphatic hydrogen or deuterium over an aromatic hydrogen or deuterium atom. The presence of aliphatic bonds also causes the PAH geometry to depart from planarity, adding stress to the molecule and resulting in weaker C-H bonds than in fully aromatic, planar PAH molecules. The lower zero-point energy of deuterium suggests that dehydrogenation is favored over dedeuteration at ISM temperatures, T < 70 K. In addition, since larger PAHs have larger heat capacities relative to small PAHs, unimolecular photodissociation favors PAHs with less than 50 carbon atoms. Theoretical estimates suggest that the expected deuterium fractionation from this method in H{\sc ii} regions is about 10$^{-5}$ \citep{Allamandola1989}. \\
\indent The low abundance or complete lack of observed deuterated features in a majority of the observed sources suggests that the conditions leading to deuterium addition in the ISM are not common. Further theoretical studies are required to assess the different scenarios in more detail. \\
\section{Conclusions}
Using AKARI, we searched for deuterated PAH emission in a sample of Galactic and extragalactic H{\sc ii} regions. We can conclude that
   \begin{enumerate}
      \item Deuterated PAHs have been observed in only six sources out of 41 in the Milky Way; this suggests that the incorporation of deuterium in PAHs is rare and highly dependent on the local conditions of the environment. The low S/N in the spectrum of LMC and SMC H{\sc ii} regions prevented us from drawing conclusions about the relation between the PAH D/H ratio and metallicity. 
      \item In our Galaxy, the average observed fractional abundance of deuterium relative to hydrogen locked up in PAHs is small, especially when compared with other interstellar molecules, such as NH$_3$. Some sources show a PAH D/H ratio upper limit as high as 0.44, or $\sim$ 0.2 if emission process is considered. Thus, while PAHs do not appear to be the sole reservoir of deuterium, they can still explain part of the variation of the galactic gas-phase D/H.
      \item Exclusion of the emission process in determining band intensities can overestimate the abundance of deuterium relative to hydrogen locked up in PAHs. The exact magnitude of this effect is found to be strongly dependent on both the size of the PAH and the characteristics of the exciting source.
      \item The PAHs are observed to be deuterium enriched relative to the galactic gas-phase abundance, since the deuterium atom is preferentially added to an aliphatic position. 
   \end{enumerate}
The upcoming James Webb Space Telescope (JWST) will offer significantly better spectral resolution in the NIR, from 1-5 $\mu$m, which will allow for better resolution of the 4.63 and 4.75 $\mu$m features, and even better constraint on the abundance of deuterium on PAHs, in the Milky Way, and in the neighboring LMC and SMC. Additionally, JWST will offer similar resolution mid-IR spectroscopy from 5-28 $\mu$m, which gives access to the intrinsically stronger C-D bending modes in addition to the C-D stretch modes. Since the C-D bending features are hard to unambiguously distinguish from the C-H bending features of PAH cations, the simultaneous detection of the C-D bending and C-D stretching features is needed for a stronger confirmation that deuterated PAHs were detected; this is a unique capability of JWST, which previous telescopes, for example, the Spitzer Space Telescope, were not able to do.

\begin{acknowledgements}
We would like to give a special thanks to the referee whose advice greatly helped the clarity of the paper. K.D.D. thanks Dr. E. Peeters for useful and stimulating discussions. This research is based on observations with AKARI, a JAXA project with the participation of ESA. Studies of interstellar PAHs at Leiden Observatory are supported through advanced European Research Council grant 246976 and a Spinoza award.        
\end{acknowledgements}

\bibliography{Ref}{}

\begin{thebibliography}{54}
\expandafter\ifx\csname natexlab\endcsname\relax\def\natexlab#1{#1}\fi

\bibitem[{{Allamandola}(1993)}]{Allamandola1993}
{Allamandola}, L.~J. 1993, in Astronomical Society of the Pacific Conference
  Series, Vol.~41, Astronomical Infrared Spectroscopy: Future Observational
  Directions, ed. S.~{Kwok}, 197

\bibitem[{{Allamandola} {et~al.}(1985){Allamandola}, {Tielens}, \&
  {Barker}}]{Allamandola1985}
{Allamandola}, L.~J., {Tielens}, A.~G.~G.~M., \& {Barker}, J.~R. 1985, \apjl,
  290, L25

\bibitem[{{Allamandola} {et~al.}(1989){Allamandola}, {Tielens}, \&
  {Barker}}]{Allamandola1989}
{Allamandola}, L.~J., {Tielens}, A.~G.~G.~M., \& {Barker}, J.~R. 1989, \apjs,
  71, 733

\bibitem[{{Bauschlicher}(1998)}]{Bauschlicher1998}
{Bauschlicher}, Jr., C.~W. 1998, \apjl, 509, L125

\bibitem[{{Bauschlicher} {et~al.}(1997){Bauschlicher}, {Langhoff}, {Sandford},
  \& {Hudgins}}]{Bauschlicher1997}
{Bauschlicher}, Jr., C.~W., {Langhoff}, S.~R., {Sandford}, S.~A., \& {Hudgins},
  D.~M. 1997, The Journal of Physical Chemistry A, 101, 2414

\bibitem[{{Bernstein} {et~al.}(2001){Bernstein}, {Dworkin}, {Sandford}, \&
  {Allamandola}}]{Bernstein2001}
{Bernstein}, M.~P., {Dworkin}, J.~P., {Sandford}, S.~A., \& {Allamandola},
  L.~J. 2001, Meteoritics and Planetary Science, 36, 351

\bibitem[{{Bertoldi} {et~al.}(2000){Bertoldi}, {Draine}, {Rosenthal},
  {Timmermann}, {Howat}, {Geballe}, {Feuchtgruber}, \&
  {Drapatz}}]{Bertoldi2000}
{Bertoldi}, F., {Draine}, B.~T., {Rosenthal}, D., {et~al.} 2000, in IAU
  Symposium, Vol. 197, From Molecular Clouds to Planetary, ed. Y.~C. {Minh} \&
  E.~F. {van Dishoeck}, 191

\bibitem[{{Boersma} {et~al.}(2014){Boersma}, {Bauschlicher}, {Ricca},
  {Mattioda}, {Cami}, {Peeters}, {S{\'a}nchez de Armas}, {Puerta Saborido},
  {Hudgins}, \& {Allamandola}}]{PAHDB}
{Boersma}, C., {Bauschlicher}, Jr., C.~W., {Ricca}, A., {et~al.} 2014, \apjs,
  211, 8

\bibitem[{{Boogert} {et~al.}(2015){Boogert}, {Gerakines}, \&
  {Whittet}}]{Boogert2015}
{Boogert}, A., {Gerakines}, P., \& {Whittet}, D. 2015, ArXiv e-prints

\bibitem[{{Buragohain} {et~al.}(2015){Buragohain}, {Pathak}, {Sarre}, {Onaka},
  \& {Sakon}}]{Buragohain2015}
{Buragohain}, M., {Pathak}, A., {Sarre}, P., {Onaka}, T., \& {Sakon}, I. 2015,
  \mnras, 454, 193

\bibitem[{{Candian} {et~al.}(2012){Candian}, {Kerr}, {Song}, {McCombie}, \&
  {Sarre}}]{Candian2012}
{Candian}, A., {Kerr}, T.~H., {Song}, I.-O., {McCombie}, J., \& {Sarre}, P.~J.
  2012, \mnras, 426, 389

\bibitem[{{Cherchneff} {et~al.}(1992){Cherchneff}, {Barker}, \&
  {Tielens}}]{Cherchneff1992}
{Cherchneff}, I., {Barker}, J.~R., \& {Tielens}, A.~G.~G.~M. 1992, \apj, 401,
  269

\bibitem[{{Dalgarno} \& {Lepp}(1984)}]{DalgarnoLepp1984}
{Dalgarno}, A. \& {Lepp}, S. 1984, \apjl, 287, L47

\bibitem[{{Demarais} {et~al.}(2014){Demarais}, {Yang}, {Snow}, \&
  {Bierbaum}}]{Demarais2014}
{Demarais}, N.~J., {Yang}, Z., {Snow}, T.~P., \& {Bierbaum}, V.~M. 2014, \apj,
  784, 25

\bibitem[{{Draine}(2004)}]{Draine2004}
{Draine}, B.~T. 2004, ArXiv Astrophysics e-prints

\bibitem[{{Epstein} {et~al.}(1976){Epstein}, {Lattimer}, \&
  {Schramm}}]{Epstein1976}
{Epstein}, R.~I., {Lattimer}, J.~M., \& {Schramm}, D.~N. 1976, \nat, 263, 198

\bibitem[{{Filipovic} {et~al.}(1998){Filipovic}, {Jones}, {White}, \&
  {Haynes}}]{Filipovic1998}
{Filipovic}, M.~D., {Jones}, P.~A., {White}, G.~L., \& {Haynes}, R.~F. 1998,
  \aaps, 130, 441

\bibitem[{{Frenklach} \& {Feigelson}(1989)}]{FrenklachFeigelson1989}
{Frenklach}, M. \& {Feigelson}, E.~D. 1989, \apj, 341, 372

\bibitem[{{Gerakines} {et~al.}(1996){Gerakines}, {Schutte}, \&
  {Ehrenfreund}}]{Gerakines1996}
{Gerakines}, P.~A., {Schutte}, W.~A., \& {Ehrenfreund}, P. 1996, \aap, 312, 289

\bibitem[{{Hilditch} {et~al.}(2005){Hilditch}, {Howarth}, \&
  {Harries}}]{Hilditch2005}
{Hilditch}, R.~W., {Howarth}, I.~D., \& {Harries}, T.~J. 2005, \mnras, 357, 304

\bibitem[{{Hudgins} {et~al.}(2004){Hudgins}, {Bauschlicher}, \&
  {Sandford}}]{Hudgins2004}
{Hudgins}, D.~M., {Bauschlicher}, Jr., C.~W., \& {Sandford}, S.~A. 2004, \apj,
  614, 770

\bibitem[{{Joblin} \& {Tielens}(2011)}]{JoblinTielens2011}
{Joblin}, C. \& {Tielens}, A.~G.~G.~M., eds. 2011, EAS Publications Series,
  Vol.~46, {PAHs and the Universe: A Symposium to Celebrate the 25th
  Anniversary of the PAH Hypothesis}

\bibitem[{{Kerridge} {et~al.}(1987){Kerridge}, {Chang}, \&
  {Shipp}}]{Kerridge1987}
{Kerridge}, J.~F., {Chang}, S., \& {Shipp}, R. 1987, \gca, 51, 2527

\bibitem[{{Kl{\ae}rke} {et~al.}(2013){Kl{\ae}rke}, {Toker}, {Rahbek},
  {Hornek{\ae}r}, \& {Andersen}}]{Klaerke2013}
{Kl{\ae}rke}, B., {Toker}, Y., {Rahbek}, D.~B., {Hornek{\ae}r}, L., \&
  {Andersen}, L.~H. 2013, \aap, 549, A84

\bibitem[{{Leger} \& {Puget}(1984)}]{LegerPuget1984}
{Leger}, A. \& {Puget}, J.~L. 1984, \aap, 137, L5

\bibitem[{{Li} \& {Draine}(2012)}]{Li2012}
{Li}, A. \& {Draine}, B.~T. 2012, \apjl, 760, L35

\bibitem[{{Linsky} {et~al.}(2006){Linsky}, {Draine}, {Moos}, {Jenkins}, {Wood},
  {Oliveira}, {Blair}, {Friedman}, {Gry}, {Knauth}, {Kruk}, {Lacour}, {Lehner},
  {Redfield}, {Shull}, {Sonneborn}, \& {Williger}}]{Linsky2006}
{Linsky}, J.~L., {Draine}, B.~T., {Moos}, H.~W., {et~al.} 2006, \apj, 647, 1106

\bibitem[{{Lis} {et~al.}(2002){Lis}, {Roueff}, {Gerin}, {Phillips}, {Coudert},
  {van der Tak}, \& {Schilke}}]{Lis2002}
{Lis}, D.~C., {Roueff}, E., {Gerin}, M., {et~al.} 2002, \apjl, 571, L55

\bibitem[{{Mart{\'{\i}}n-Hern{\'a}ndez}
  {et~al.}(2002{\natexlab{a}}){Mart{\'{\i}}n-Hern{\'a}ndez}, {Peeters},
  {Morisset}, {Tielens}, {Cox}, {Roelfsema}, {Baluteau}, {Schaerer}, {Mathis},
  {Damour}, {Churchwell}, \& {Kessler}}]{MartinHernandez2002a}
{Mart{\'{\i}}n-Hern{\'a}ndez}, N.~L., {Peeters}, E., {Morisset}, C., {et~al.}
  2002{\natexlab{a}}, \aap, 381, 606

\bibitem[{{Mart{\'{\i}}n-Hern{\'a}ndez}
  {et~al.}(2002{\natexlab{b}}){Mart{\'{\i}}n-Hern{\'a}ndez}, {Vermeij},
  {Tielens}, {van der Hulst}, \& {Peeters}}]{MartinHernandez2002b}
{Mart{\'{\i}}n-Hern{\'a}ndez}, N.~L., {Vermeij}, R., {Tielens}, A.~G.~G.~M.,
  {van der Hulst}, J.~M., \& {Peeters}, E. 2002{\natexlab{b}}, \aap, 389, 286

\bibitem[{{Mori} {et~al.}(2014){Mori}, {Onaka}, {Sakon}, {Ishihara},
  {Shimonishi}, {Ohsawa}, \& {Bell}}]{Mori2014}
{Mori}, T.~I., {Onaka}, T., {Sakon}, I., {et~al.} 2014, \apj, 784, 53

\bibitem[{{Onaka} {et~al.}(2009){Onaka}, {Lorente}, \& {Ita}}]{Onaka2009}
{Onaka}, T., {Lorente}, R., \& {Ita}, Y. 2009, {IRC Data Users's Manual for
  Phase 3 ver 1.1}

\bibitem[{{Onaka} {et~al.}(2007){Onaka}, {Matsuhara}, {Wada}, {Fujishiro},
  {Fujiwara}, {Ishigaki}, {Ishihara}, {Ita}, {Kataza}, {Kim}, {Matsumoto},
  {Murakami}, {Ohyama}, {Oyabu}, {Sakon}, {Tanab{\'e}}, {Takagi}, {Uemizu},
  {Ueno}, {Usui}, {Watarai}, {Cohen}, {Enya}, {Ootsubo}, {Pearson}, {Takeyama},
  {Yamamuro}, \& {Ikeda}}]{Onaka2007}
{Onaka}, T., {Matsuhara}, H., {Wada}, T., {et~al.} 2007, \pasj, 59, 401

\bibitem[{{Onaka} {et~al.}(2014){Onaka}, {Mori}, {Sakon}, {Ohsawa}, {Kaneda},
  {Okada}, \& {Tanaka}}]{Onaka2014}
{Onaka}, T., {Mori}, T.~I., {Sakon}, I., {et~al.} 2014, \apj, 780, 114

\bibitem[{{Paladini} {et~al.}(2003){Paladini}, {Burigana}, {Davies}, {Maino},
  {Bersanelli}, {Cappellini}, {Platania}, \& {Smoot}}]{Paladini2003}
{Paladini}, R., {Burigana}, C., {Davies}, R.~D., {et~al.} 2003, \aap, 397, 213

\bibitem[{{Parise} {et~al.}(2004){Parise}, {Castets}, {Herbst}, {Caux},
  {Ceccarelli}, {Mukhopadhyay}, \& {Tielens}}]{Parise2004}
{Parise}, B., {Castets}, A., {Herbst}, E., {et~al.} 2004, \aap, 416, 159

\bibitem[{{Peeters} {et~al.}(2004){Peeters}, {Allamandola}, {Bauschlicher},
  {Hudgins}, {Sandford}, \& {Tielens}}]{Peeters2004}
{Peeters}, E., {Allamandola}, L.~J., {Bauschlicher}, Jr., C.~W., {et~al.} 2004,
  \apj, 604, 252

\bibitem[{{Pietrzy{\'n}ski} {et~al.}(2013){Pietrzy{\'n}ski}, {Graczyk},
  {Gieren}, {Thompson}, {Pilecki}, {Udalski}, {Soszy{\'n}ski}, {Koz{\l}owski},
  {Konorski}, {Suchomska}, {Bono}, {Moroni}, {Villanova}, {Nardetto},
  {Bresolin}, {Kudritzki}, {Storm}, {Gallenne}, {Smolec}, {Minniti}, {Kubiak},
  {Szyma{\'n}ski}, {Poleski}, {Wyrzykowski}, {Ulaczyk}, {Pietrukowicz},
  {G{\'o}rski}, \& {Karczmarek}}]{Pietrzynski2013}
{Pietrzy{\'n}ski}, G., {Graczyk}, D., {Gieren}, W., {et~al.} 2013, \nat, 495,
  76

\bibitem[{{Pilleri} {et~al.}(2015){Pilleri}, {Joblin}, {Boulanger}, \&
  {Onaka}}]{Pilleri2015}
{Pilleri}, P., {Joblin}, C., {Boulanger}, F., \& {Onaka}, T. 2015, \aap, 577,
  A16

\bibitem[{{Rasmussen} {et~al.}(2011){Rasmussen}, {Henkelman}, \&
  {Hammer}}]{Rasmussen2011}
{Rasmussen}, J.~A., {Henkelman}, G., \& {Hammer}, B. 2011, \jcp, 134, 164703

\bibitem[{{Rauls} \& {Hornek{\ae}r}(2008)}]{Rauls2008}
{Rauls}, E. \& {Hornek{\ae}r}, L. 2008, \apj, 679, 531

\bibitem[{{Ricca} {et~al.}(2007){Ricca}, {Bakes}, \&
  {Bauschlicher}}]{Ricca2007}
{Ricca}, A., {Bakes}, E.~L.~O., \& {Bauschlicher}, Jr., C.~W. 2007, \apj, 659,
  858

\bibitem[{{Roberts} {et~al.}(2002){Roberts}, {Fuller}, {Millar}, {Hatchell}, \&
  {Buckle}}]{Roberts2002}
{Roberts}, H., {Fuller}, G.~A., {Millar}, T.~J., {Hatchell}, J., \& {Buckle},
  J.~V. 2002, \planss, 50, 1173

\bibitem[{{Roberts} {et~al.}(2003){Roberts}, {Herbst}, \&
  {Millar}}]{Roberts2003}
{Roberts}, H., {Herbst}, E., \& {Millar}, T.~J. 2003, \apjl, 591, L41

\bibitem[{{Sandford} {et~al.}(2000){Sandford}, {Bernstein}, {Allamandola},
  {Gillette}, \& {Zare}}]{Sandford2000}
{Sandford}, S.~A., {Bernstein}, M.~P., {Allamandola}, L.~J., {Gillette}, J.~S.,
  \& {Zare}, R.~N. 2000, \apj, 538, 691

\bibitem[{{Sandford} {et~al.}(2001){Sandford}, {Bernstein}, \&
  {Dworkin}}]{Sandford2001}
{Sandford}, S.~A., {Bernstein}, M.~P., \& {Dworkin}, J.~P. 2001, Meteoritics
  and Planetary Science, 36, 1117

\bibitem[{{Sandford} {et~al.}(2013){Sandford}, {Bernstein}, \&
  {Materese}}]{Sandford2013}
{Sandford}, S.~A., {Bernstein}, M.~P., \& {Materese}, C.~K. 2013, \apjs, 205, 8

\bibitem[{{Schutte} {et~al.}(1993){Schutte}, {Tielens}, \&
  {Allamandola}}]{Schutte1993}
{Schutte}, W.~A., {Tielens}, A.~G.~G.~M., \& {Allamandola}, L.~J. 1993, \apj,
  415, 397

\bibitem[{{Storey} \& {Hummer}(1995)}]{StoreyHummer1995}
{Storey}, P.~J. \& {Hummer}, D.~G. 1995, \mnras, 272, 41

\bibitem[{{Thrower} {et~al.}(2012){Thrower}, {J{\o}rgensen}, {Friis},
  {Baouche}, {Mennella}, {Luntz}, {Andersen}, {Hammer}, \&
  {Hornek{\ae}r}}]{Thrower2012}
{Thrower}, J.~D., {J{\o}rgensen}, B., {Friis}, E.~E., {et~al.} 2012, \apj, 752,
  3

\bibitem[{{Tielens}(1992)}]{Tielens1992}
{Tielens}, A.~G.~G.~M. 1992, in IAU Symposium, Vol. 150, Astrochemistry of
  Cosmic Phenomena, ed. P.~D. {Singh}, 91

\bibitem[{{Tielens}(2008)}]{Tielens2008}
{Tielens}, A.~G.~G.~M. 2008, \araa, 46, 289

\bibitem[{{van der Tak} {et~al.}(2002){van der Tak}, {Schilke}, {M{\"u}ller},
  {Lis}, {Phillips}, {Gerin}, \& {Roueff}}]{vanderTak2002}
{van der Tak}, F.~F.~S., {Schilke}, P., {M{\"u}ller}, H.~S.~P., {et~al.} 2002,
  \aap, 388, L53

\bibitem[{{Vidal-Madjar} {et~al.}(1998){Vidal-Madjar}, {Lemoine}, {Ferlet},
  {Hebrard}, {Koester}, {Audouze}, {Casse}, {Vangioni-Flam}, \&
  {Webb}}]{VidalMadjar1998}
{Vidal-Madjar}, A., {Lemoine}, M., {Ferlet}, R., {et~al.} 1998, \aap, 338, 694

\end{thebibliography}
\bibliographystyle{aa}

\end{document}